# Fluctuation magnetoconductivity in pristine and proton-irradiated $Ca_{8.5}La_{1.5}(Pt_3As_8)(Fe_2As_2)_5$ single crystals

D. Ahmad, Y. I. Seo, W. J. Choi, and Yong Seung Kwon[*]

Department of Emerging Materials Science, DGIST, Daegu, 42988, Republic of Korea

## Abstract

The influence of the proton irradiation on the thermally fluctuation–induced conductivity in $Ca_{8.5}La_{1.5}(Pt_3As_8)(Fe_2As_2)_5$ single crystal was investigated. The in-plane magnetoconductivity was measured up to $\mu_0 H$=13 T. It is observed that the $T_c$ was suppressed up to 30.3 from 32.5 K when the proton is irradiated whereas the amplitude of the fluctuation effect is the almost same in both samples. The results were analyzed by the Ullah and Dorsey scaling functions based on the Hartree approximation to the Ginzburg-Landau free energy, including the critical-fluctuation regime. In the pristine sample, the 3D-2D crossover in the fluctuation effect was observed near the $T_c$. When the 3D-2D occurs, it is newly found that there is a regime simultaneously described by 2D and 3D fluctuation behaviors. Meanwhile, the proton-irradiated sample showed the 3D fluctuation.

[*]Corresponding Author: yskwon@dgist.ac.kr

## Introduction

The recent discovery of superconductivity in iron-pnictides [1], has led a new direction for the condensed matter physicists for investigating the superconducting phenomena and comparison with other high temperature superconductors. The iron-pnictides superconductors show many similarities with cuprate superconductors, for instance, the two dimensional layered-structure [2], the properties dramatically changing with doping, the proximity to an antiferromagnetic phase [3], the relatively large ratio ($\Delta/k_B T_c$) between the superconducting (SC) gap large ratio between the superconducting (SC) gap $\Delta$ [4], and the small superfluid density [5]. However, different from cuprates, the iron-pnictides superconductors show the metallic properties in undoped state (insulator in cuprates) [2], the multiband character of superconductivity due to several Fermi surfaces and relatively small anisotropy ratio [6]. Moreover, since the electron-phonon coupling cannot theoretically explain the high values of $T_c$, it has been suggested that the pairing glue is formed by spin fluctuations exchanged between electrons in different bands [7].

Thermal fluctuations in high-$T_c$ cuprate superconductors are significant owing to the high-transition temperature. These fluctuations are enhanced by the short coherence length and the low dimensionality resulting from the layered structure and the high anisotropy. By analyzing the fluctuation effects on the physical properties such as conductivity, magnetization, and thermoelectricity, important information about superconductivity can be deduced from the theoretical predictions studied well [8-11].

The magnitude of thermal fluctuation is quantified by the Ginzburg number, $Gi = 10^{-9}(\kappa^4 T_c[K]\gamma^2/H_{c2}(0)[Oe])$, where $\kappa = \lambda_0/\xi_0$ is the Ginzburg-Landau parameter, $\gamma = \xi_{ab}/\xi_c$ is the anisotropy ratio. The Ginzburg number is $10^{-9}$–$10^{-6}$ for conventional low-$T_c$ superconductors, while is $10^{-2}$–$10^{-1}$ for high-$T_c$ cuprates [12]. Strong thermal fluctuations give a rich variety of vortex phases and vortex dynamics in high-$T_c$ superconductors. Even though the $T_c$ is lower than that of high-$T_c$ cuprates, a quite pronounced fluctuation effect is also observed in intermetallic superconductors, for example in YNi$_2$B$_2$C [13] and in the MgB$_2$ superconductors with $Gi$~$10^{-6}$ [14].

In iron-based superconductors, $Gi$ is of the order of $10^{-3}$ for SmFeAs(O$_{1-x}$F$_x$) [15], and $10^{-5}$ for Co-doped BaFe$_2$As$_2$ [16] and LiFeAs single crystals [17]. In our Ca$_{8.5}$La$_{1.5}$(Pt$_3$As$_8$)(Fe$_2$As$_2$)$_5$, $Gi$ is of the order of $10^{-4}$, which is considerably higher than that of YNi$_2$B$_2$C, and thus the fluctuation-induced conductivity is expected to be observed in Ca$_{8.5}$La$_{1.5}$(Pt$_3$As$_8$)(Fe$_2$As$_2$)$_5$ but has been not yet studied.

In this paper, we apply fluctuation and proton irradiation techniques more profoundly to investigate the fundamental parameters for superconducting fluctuation in Ca$_{8.5}$La$_{1.5}$(Pt$_3$As$_8$)(Fe$_2$As$_2$)$_5$ single crystal. The proton irradiation is a useful technique to introduce effective pinning centers for the enhancement of the critical current density. In particular, we will study the fluctuation-induced in-plane conductivity under magnetic fields up to 13 T. The results were analyzed by the Ullah and Dorsey scaling functions based on the Hartree approximation to the Ginzburg-Landau free energy, including the critical-fluctuation regime [8,9].

**Experiments**

We grew the single crystal of $Ca_{8.5}La_{1.5}(Pt_3As_8)(Fe_2As_2)_5$ using a Bridgman method with sealed molybdenum and boron nitride (BN) crucibles in order to prevent the evaporation of arsenic. In the first step, precursors of CaAs, LaAs and FeAs were produced in evacuated quartz ampoules at 550, 800 and 1050°C, respectively. In the second step, the precursors and platinum were mixed together and put in the BN crucible and then the BN crucible was put in the Mo-crucible welded with an *arc* welding system. The entire assembly was slowly heated up to 1350°C in vacuum furnace with a tungsten meshed heater and then moved in a downward direction with 2 mm/hr rate. X-ray diffraction (XRD) of the powder made from the sample was performed with Cu-K$\alpha$ radiation. The Rietveld refinement results for the XRD data showed that $Ca_{0.85}La_{0.15}(Pt_3As_8)(Fe_2As_2)_5$ crystallized in a triclinic $SrZnBi_2$ structure with lattice parameters of *a* = 8.7621, 8.7593 and *c* = 10.7508 Å and the angle of $\alpha$=94.644, $\beta$=104.493 and $\gamma$=90.258°, which is similar to the reported results [18].

Proton irradiation was performed at 300 K using an MC-50 cyclotron installed on KIRAM (Korea institute of radiological & medical sciences). The total dose is $5 \times 10^{15}$ p/cm$^2$. To prevent the temperature rise we used the weak flux density rate of $1.964 \times 10^{11}$ p/cm$^2$·s and the copper block, on which samples is attached, flowed by water. During the irradiation, the temperature of copper block keeps 300 K.

The in-plane resistivity measurements using a four-point collinear probe were performed using a 16T Oxford superconducting magnet system. The temperature dependence of the transverse in-plane resistivity was measured at $\mu_0H$=0, 1, 2, 3, 5, 7, 9, 11 and 13 T for *H*//c axis.

**Results and Discussion**

Figure 1 (a) and (b) show the temperature dependence of the in-plane resistivity of pristine and proton-irradiated $Ca_{8.5}La_{1.5}(Pt_3As_8)(Fe_2As_2)_5$ single crystals at various applied magnetic fields. As it is clear from Fig. 1 the normal state resistivity in the pristine sample is of a half-amplitude of that in the proton-irradiated sample. When the transition temperature is roughly determined from the transition midpoint at zero applied field, the $T_c$ value is 32.5 K for the pristine sample, which is in agreement with the previously reported single crystal, and is 30.3 K. Furthermore, one can see that by increasing the applied magnetic field, $T_c$ is shifted to lower temperatures and the transition width becomes broader in both samples. The broadness is well known to be due to the thermal fluctuation of Cooper pairs originating from the high-transition temperature.

As a result of thermal fluctuations, there is a finite probability of Cooper-pair formation above $T_c$. This gives rise to an excess conductivity (EC) Δσ for $T > T_c$. This effect has been studied by Aslamazov-Larkin (AL) [19] and Maki-Thompson (MK) [20, 21]. It is well known that the MK term is much less than the AL term and plays a minor role near $T_c$. Hence, the experimental conductivity could be described as σ =σ$_n$+Δσ$_{AL}$. Here, the leading contributions to EC take two forms due to the coupling strength between the conducting planes:

$$\Delta\sigma_{3D} = \frac{e^2}{32\hbar} \frac{1}{\xi_c(0)} \epsilon^{-1/2}, \quad (1)$$

$$\Delta\sigma_{2D} = \frac{e^2}{16\hbar} \frac{1}{d} \epsilon^{-1}, \quad (2)$$

where $\epsilon$, $\xi_c(0)$ and $d$ are the reduced temperature $(T - T_c^{MF})/T_c^{MF}$, the coherence length along the direction perpendicular to the layers and a characteristic length of the two dimensional system. Here, $T_c^{MF}$ is the mean-field transition temperature.

To probe the fluctuation dimensionality, the EC in zero magnetic field due to finite Cooper-pair formation above $T_c$ was analyzed by AL terms. Δσ is defined as 1/ρ(T)−1/ρ$_n$(T), where ρ(T) is the actual measured resistivity and ρ$_n$(T) is the extrapolated normal state resistivity.

In view of the linear temperature dependence of $d\rho(T)/dT$ (an example for μ$_0$H=0 T is presented in Fig. 2(a), the background contributions were estimated by fitting a quadratic polynomial to the measured ρ(T) from ~2$T_c$ down to $T_{onset}$, below which fluctuation effects are measurable. The fitted normal state resistivity is plotted in Fig. 2(b). $T_c^{MF}$ was decided from the temperature on the peak in the first derivative and zero in the second derivative of $\rho(T)$ (an example for μ$_0$H=0 T is presented in Fig. 2(a).

Figure 3 shows the reduced temperature dependence of the excess conductivity $\Delta\sigma(\epsilon)$ at μ$_0$H=0 T. As shown in Fig. 3(a), the 3D behavior of $\Delta\sigma \propto \epsilon^{-0.5}$ was observed in the low-temperature region for the pristine sample. Using the eq. (1), the coherence length along the direction perpendicular to the layers is estimated to be $\xi_c(0)$=0.527 nm. 2$\xi_c(0)$ is comparable to the lattice constant of *c*-axis, which is consistent with 3D fluctuations result. This 3D behavior changes into 2D behavior of $\Delta\sigma \propto \epsilon^{-1}$ with increasing temperature, which is similar in the case of cuprates high $T_c$ superconductors [22-25]. Using the eq. (2), a characteristic length is evaluated to be $d$=4.37 nm, which is comparable to the 4$c$. The $d$-value was seemingly evaluated largely because of the very weak superconducting fluctuations shown in considerably higher temperature regime than $T_c$. There are a few reports on FC for iron-based superconductors but the temperature-dependent 2D-3D crossover behaviors have not yet been reported, except for the result of the F-doped SmFeAsO [26] and LiFeAs [17].

In the presence of the magnetic field, the excess conductivity is analyzed including the critical fluctuation regime in the framework of 2D or 3D scaling behaviors. The critical fluctuation effects are theoretically calculated by considering the fourth-order term in GL free energy, within a self-consistent Hartree approximation by Ullah and Dorsey (UD scaling) [8, 9]. The scaling forms are given by

$$\Delta\sigma_{3D} = \left(\frac{T^2}{H}\right)^{1/3} \widetilde{F_{3D}} \left(A \frac{T-T_c(H)}{(TH)^{2/3}}\right), \qquad (3)$$

$$\Delta\sigma_{2D} = \left(\frac{T}{H}\right)^{1/2} \widetilde{F_{2D}} \left(B \frac{T-T_c(H)}{(TH)^{1/2}}\right), \qquad (4)$$

where A and B are field and temperature independent constants characterizing the materials, and $\widetilde{F_{3D}}$ and $\widetilde{F_{2D}}$3D are unspecified scaling functions. In UD scaling, the critical temperature $T_c(H)$ used the mean field transition temperature $T_c^{MF}$, obtained above, fine-tuned with uncertainty of ±0.5%. In Fig. 4(a) and (b), the scaled magnetoconductivity of Ca$_{8.5}$La$_{1.5}$(Pt$_3$As$_8$)(Fe$_2$As$_2$)$_5$ single crystal is given by the $\Delta\sigma(H/T^2)^{1/3}$ vs $[T-T_c(H)]/(TH)^{2/3}$ of the 3D case and $\Delta\sigma(H/T)^{1/2}$ vs $[T-T_c(H)]/(TH)^{1/2}$ of the 2D case, respectively. Excellent scaling behavior of fluctuations for the 3D case is obtained in the low temperature regime satisfying the condition of $[T-T_c(H)]/(TH)^{2/3}$< 0.026 K$^{1/3}$ T$^{-2/3}$ but in higher temperature regime poor scaling behavior was observed (see the inset of Fig. 4 (a). On the other hand, the 2D scaling plot gives excellent result above $T_c$ but poor result below $T_c$ (see the inset of Fig. 4 (b)). The $T_c(H)$ was used same values in 2 and 3D scaling. The used $T_c(H)$ and border lines of 3D and 2D are plotted in Fig. 5. The $T_c(H)$ is similar to $T_c^{MF}(H)$, which is obtain above, within error bars and decreases linearly as the rate of $\partial H_{c2}/\partial T = -4.56$ T/K with increasing the magnetic field above μ$_0$H=3 T. From the Ginzburg-Landau theory, the shift is described by

$$\ln\left(\frac{T_c(0)}{T_c(H)}\right) = \frac{H}{T_c(0)} \left|\frac{\partial H_{c2}}{\partial T}\right|^{-1}_{T_c(0)}. \qquad (5)$$

Using the $\partial H_{c2}/\partial T = -4.56$ T/K, we obtained $T_c$(0)=31.23 K, which is significantly different with $T_c^{MF}(0)$=32.49 K. This difference attributes to multiband properties because the $H_{c2}$(T) curve is often bent in low magnetic field region of the multiband superconductors such as FeAs-based superconductors[27]. As shown in Fig. 5, the superconducting fluctuation changes from 3D to 2D behavior, which is consistent with AL-analysis at $\mu_0H$=0, mentioned above, and has been in cuprates[22-25], F-doped SmFeAsO[26] and LiFeAs[17] already. When the dimensional crossover occurs, the overlap region was observed in $Ca_{8.5}La_{1.5}(Pt_3As_8)(Fe_2As_2)_5$ single crystal and has been not yet reported to date.

The AL analysis for the same sample at $\mu_0H$=0 T after proton irradiation is presented in Fig. 6. The 3D behavior of $\Delta\sigma \propto \epsilon^{-0.5}$ was observed up to a higher temperature compared with the pristine sample. It is known that the deviation above the temperature is due to an overestimation for the energy of the fluctuation modes in AL theory [28-32]. Using the eq. (1), the coherence length is estimated to be $\xi_c(0)$=0.993 nm. $2\xi_c(0)$ is considerably larger than the lattice constant of *c*-axis, which is consistent with 3D fluctuation result.

In UD scaling for the proton-irradiated sample, the critical temperature $T_c(H)$ also used the mean field transition temperature $T_c^{MF}$ fine-tuned with uncertainty of ±0.5%. In Fig. 6 (a) and (b), the scaled magnetoconductivity is given by the $\Delta\sigma(H/T^2)^{1/3}$ vs $[T - T_c(H)]/(TH)^{2/3}$ of the 3D case and $\Delta\sigma(H/T)^{1/2}$ vs $[T - T_c(H)]/(TH)^{1/2}$ of the 2D case, respectively. Excellent scaling behavior of fluctuations is obtained for the 3D case, while the 2D scaling plot gives poor result, which is different with the result of pristine sample. The $T_c(H)$ was used same values in 2D and 3D scaling. The used $T_c(H)$ is plotted in the right inset of Fig. 6(a) and decreases linearly as the rate of $\partial H_{c2}/\partial T = -3.24$ T/K with increase the magnetic field above $\mu_0H$=3T. Using the value of $\partial H_{c2}/\partial T$, we obtained $T_c(0)$=29.59 K from the Ginzburg-Landau theory (eq. (5)), which is significantly different with $T_c^{MF}(0)$=30.30 K. This difference attributes to multiband properties similar to the result of the pristine sample.

## Conclusions

Superconducting $Ca_{8.5}La_{1.5}(Pt_3As_8)(Fe_2As_2)_5$ single crystal was grown by a Bridgman method. The effect of proton irradiation on the superconducting fluctuation effects including the critical-fluctuation regime was analyzed on the pristine and proton-irradiated samples in terms of the Ullah and Dorsey scaling functions based on the Hartree approximation to the Ginzburg-Landau free energy. The scaling behavior of the pristine sample showed the 3D-2D crossover near $T_c$. When the crossover occurs, temperature regime overlapping of 2D and 3D fluctuation effects was found near $T_c$. On the other hand, the proton-irradiated sample was well described with the 3D theoretical predictions.

# References

[1] Kamihara Y, Watanabe T, Hirano M, and Hosono H 2008 *J. Am. Chem. Soc.* **130** 3296.

[2] Hosono H, Tanabe K, Takayama-Muromachi E, Kageyama H, Yamanaka S, Kumakura H, Nohara M, Hiramatsu H, Fujitsu S 2015 *Science and Technology of Advanced Materials.* **16** 033503.

[3] Zhao J, Huang Q, Cruz C. D. L, Li S, Lynn J W, Chen Y, Green M A, Chen G F, Li G, Li Z, Luo J L, Wang N L and Dai P 2008 *Nature Materials*. 7 953 ; Chen H, Ren Y, Qiu Y, Bao W, Liu R H, Wu G, Wu T, Xie Y L, Wang X F, Huang Q and Chen X H 2009 *Europhys. Lett*. **85**, 17006; Canfield P C and S. L. Budko S L 2010 *Annu. Rev. Condens. Matter Phys*. **1** 27.

[4] Popovich P, Boris A V, Dolgov O V, Golubov A A, Sun D L, Lin C T, Kremer R K and Keimer B 2010 *Phys. Rev. Lett*. 105 027003.

[5] Gang M, Xi-Yu Z, Lei F, Lei S, Cong R and Hai-Hu W 2008 *Chin. Phys. Lett.* **25** 2221.

[6] Singh D J and Du M H 2008 *Phys. Rev Lett.* **100** 237003; Singh D J 2008 *Phys. Rev B* **78** 094511.

[7] Boeri L, Dolgov O V and Golubov A A 2008 *Phys. Rev. Lett.* **101** 026403.

[8] Ullah S and Dorsey A T 1991 *Phys. Rev. B* **44** 262.

[9] Ullah S and Dorsey A T 1991 *Phys. Rev. Lett.* **65** 2066.

[10] Ikeda R, Ohmi T and Tsuento T 1991 *Phys. Rev. Lett.* **67** 3874.

[11] Tesanovic Z, Xing L, Bulaevskii L, Li Q, and Suenaga M 1992 *Phys. Rev. Lett.* **69** 3563.

[1*2*] Masui T, Lee S and Tajima S 2003 *Physica C* **383** 299.

[13] Mun M O, Lee S I and Lee W C 1997 Phys. Rev. B **56** 14668.

[14] Pribulova Z, Klein T, Kacmarcik J, Marcenat C, Konczykowski M, Bud'ko S L, Tillman M and Canfield P C 2009 *Phys. Rev. B* **79** 020508.

[15] Pallecchi I, Fanciulli C, Tropeano M, Palenzona A, Ferretti M, Malagoli A, Martinelli A, Sheikin I, Putti M and Ferdeghini C 2009 *Phys. Rev. B* **79** 104515.

[16] Kim S H, Choi C H, Jung M H, Yoon J B, Jo Y H, Wang X F, Chen X H, Wang X L, Lee S I and Choi K Y, 2010 *J. Appl. Phys*. **108** 063916.

[17] Song Y J, Kang B, Rhee J S and Y. S. Kwon 2012 *Europhys. Lett*. **97** 47003.
[18] Stürzer T, Derondeau G and Johrendt D 2012 *Phy. Rev. B* **86** 060516(R).

[19] Aslamazov L G and Larkin A I 1968 *Phys. Lett. A*. **26** 238.

[20] Maki K 1968 *Prog. Theor. Phys.* **39** 897 ; 1968 **40** 193.

[21] Thompson R S 1970 *Phys. Rev. B* **1** 327.

[22] Safar H, Gammel P L, Bishop D J, Mitzi D B and Kapitulnik A 1992 *Phys. Rev. Lett.* **68** 2672.

[23] Hassan N, Shabbir B and Khan N A 2009 *J. Appl. Phys.* **105** 083926.

[24] Zavgorodnii A A, Vovk R V, Obolenskii M A and Samoilov A V 2010 *Low Temp. Phys.* **36** 115.

[25] Ausloos M and Ch. Laurent Ch 1988 *Phys. Rev. B* **37** 611.

[26] Liu S L, Gang B and Wang H 2010 *J. Supercond. Nov. Magn*. **23** 1563.

[27] Hunte F, Jaroszynski J, Gurevich A, Larbalestier D C, Jin R, Sefat A S, McGuire M A, Sales B C, Christen D K and Mandrus D 2008 *Nature* **453** 903.

[28] Johnson W L and Tsuei C C 1976 *Phys. Rev. B* **13**, 4827.

[29] Johnson W L, Tsuei C C and Chaudhari P, *Phys. Rev. B* **17**, 2884 (1978).

[30] Carballeira C, Mosqueira J, Revcolevschi A and Vidal F 2000 *Phys. Rev. Lett.* **84**, 3157.

[31] Carballeira C, Mosqueira J, Revcolevschi A and Vidal F 2003 *Physica C* **384** 185.

[32] Gollub J P, Beasley M R, Callarotti R and Tinkham M 1973 *Phys. Rev. B* **7** 3039.

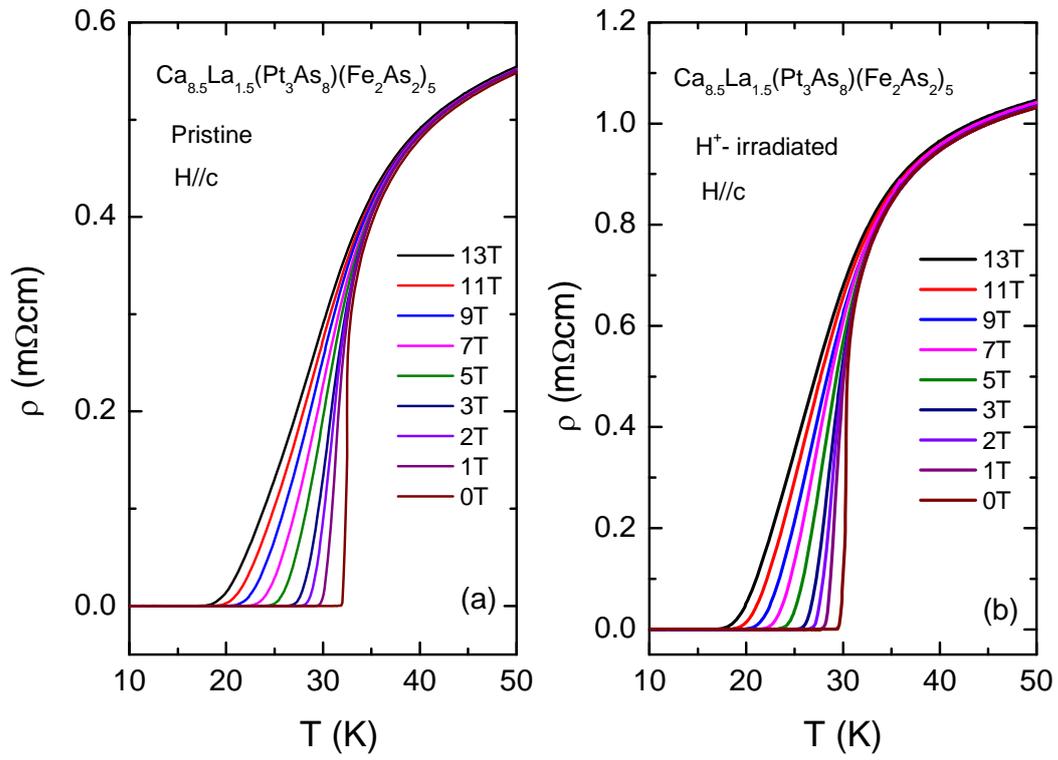

**Fig. 1.** Temperature dependence of in-plane resistivities at various fields in the pristine and the proton-irradiated $Ca_{8.5}La_{1.5}(Pt_3As_8)(Fe_2As_2)_5$ single crystals for H//*c*-axis.

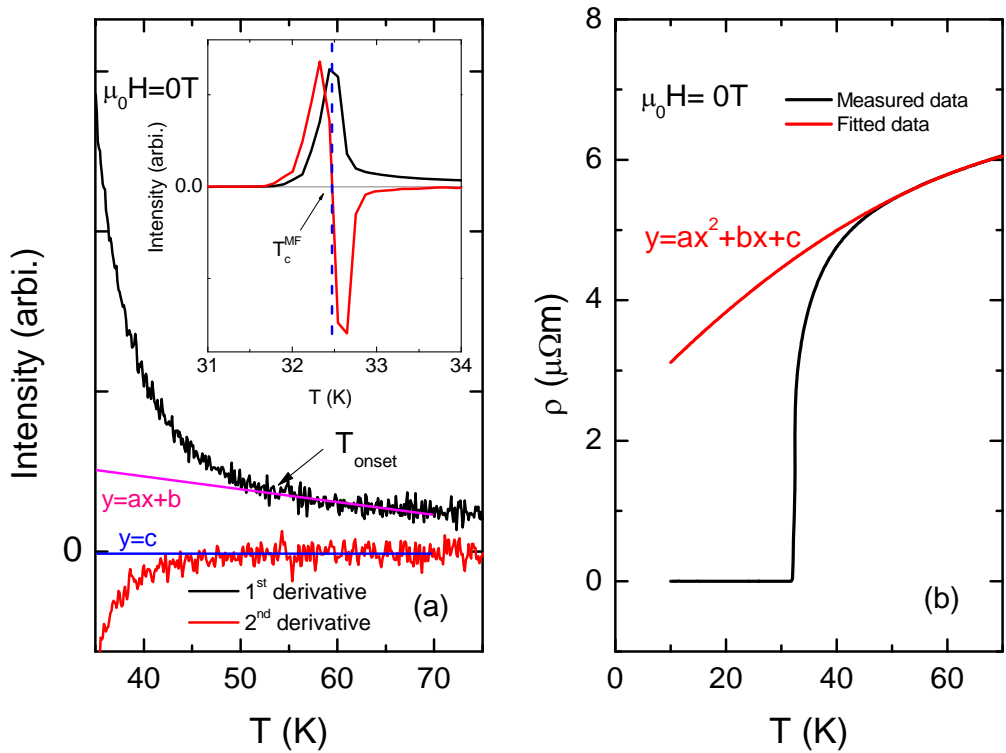

**Fig. 2.** (a) Temperature dependence of the first (black line) and second (red line) derivative of resistivity at $\mu_0 H=0$ T with respect with temperature. The pink-colored solid line is the fit of degree 1 polynomial and the blue-colored solid line is the fit of constant value. Inset shows the first (black line) and second (red line) derivative of resistivity near transition temperature. (b) Temperature dependence of the measured resistivity (black line) and the data fitted by a degree 2 polynomial (red line) at $\mu_0 H=0$ T.

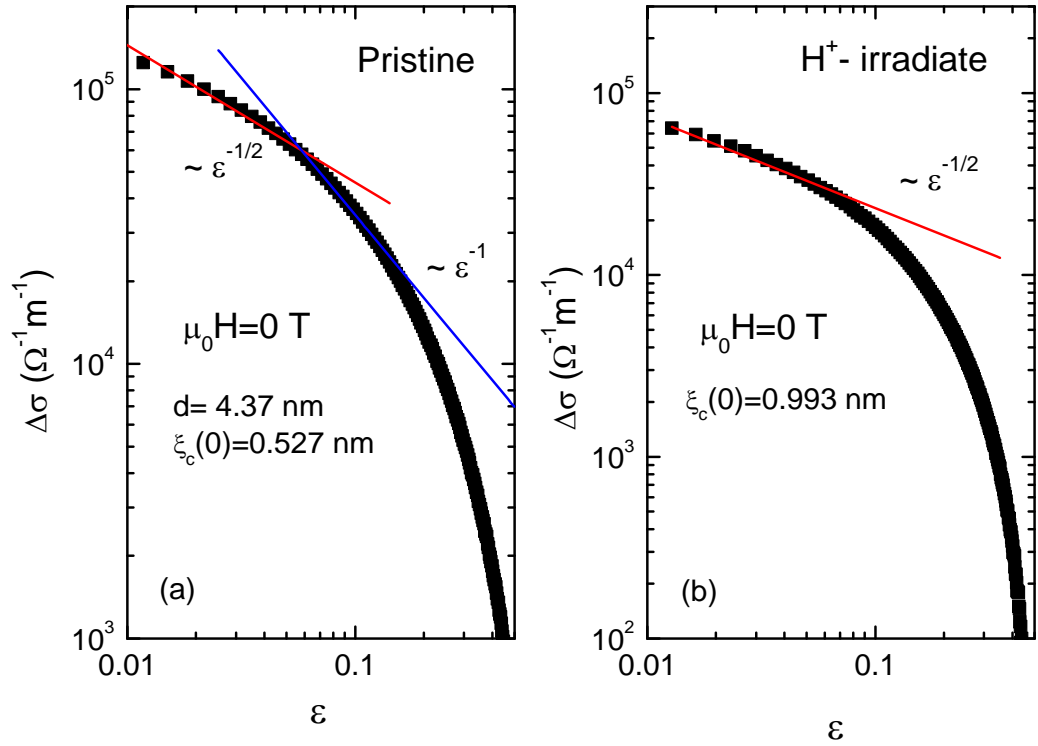

**Fig. 3.** Superconducting fluctuation conductivity as a function of ε in a log-log scale for pristine (a) and proton-irradiated (b) samples. Here $\varepsilon = (T - T_c^{MF})/T_c^{MF}$ is the reduced temperature. The solid red-colored line represents the 3D-AL behavior, and the solid blue-colored line is the 2D AL behavior.

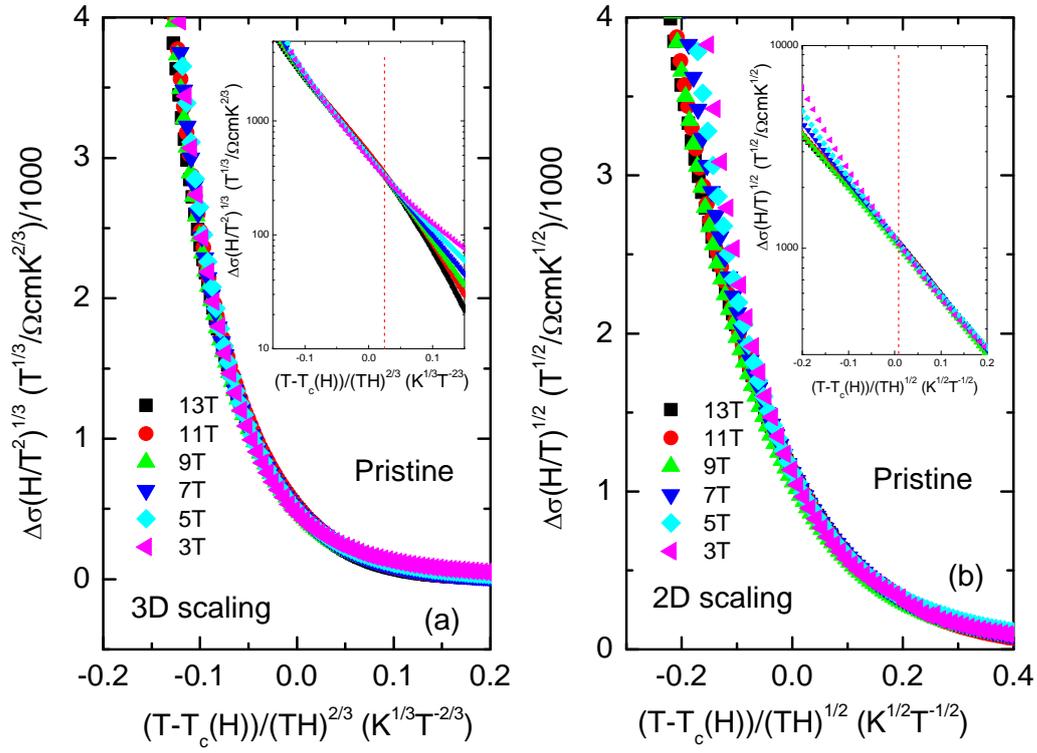

**Fig. 4.** In the main panels, the plots of the quantity $\Delta\sigma(H/T^2)^{1/3}$ vs $[T - T_c(H)]/(TH)^{2/3}$ (a) and $\Delta\sigma(H/T)^{1/2}$ vs $[T - T_c(H)]/(TH)^{1/2}$ (b) expected for the 3D-UD and 2D-UD cases are shown for the pristine $Ca_{8.5}La_{1.5}(Pt_3As_8)(Fe_2As_2)_5$. In the insets, the same plots are displayed in semi-logarithmic scale.

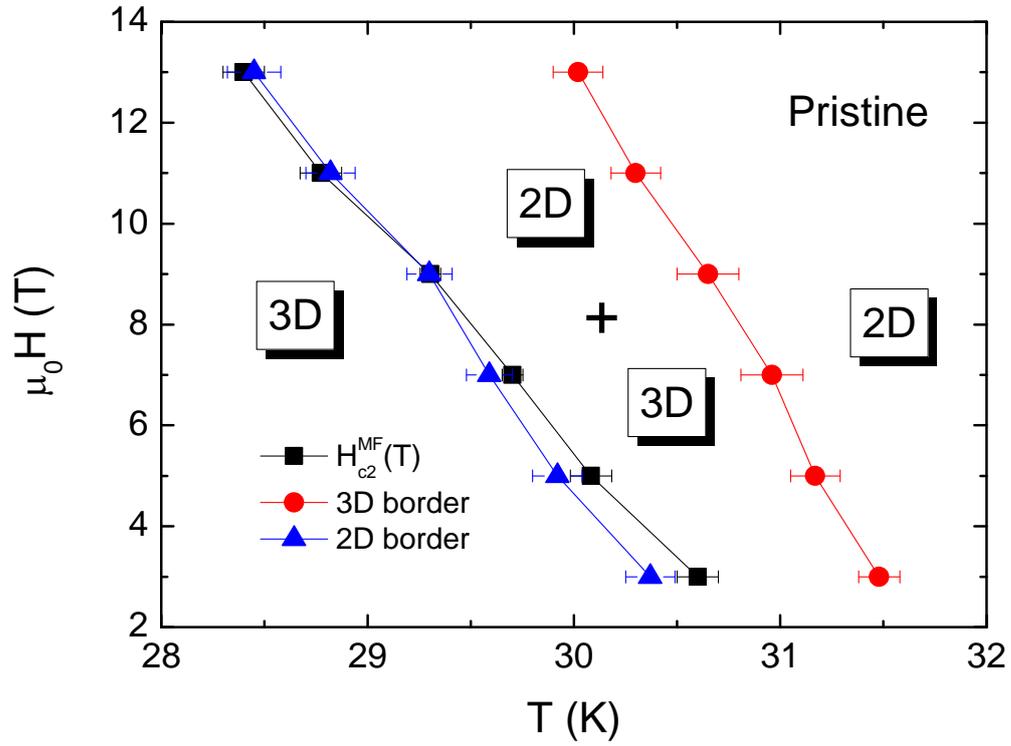

**Fig. 5.** Fluctuation dimension phase diagram of pristine $Ca_{8.5}La_{1.5}(Pt_3As_8)(Fe_2As_2)_5$. The upper critical fields $H_{c2}^{MF}$ were estimated from the mean field transition temperatures. In the "3D", "2D" and "2D+3D" zones, respectively, superconducting fluctuation conductivity is well described by three dimensional UD, two dimensional UD and both dimensional UD scalings.

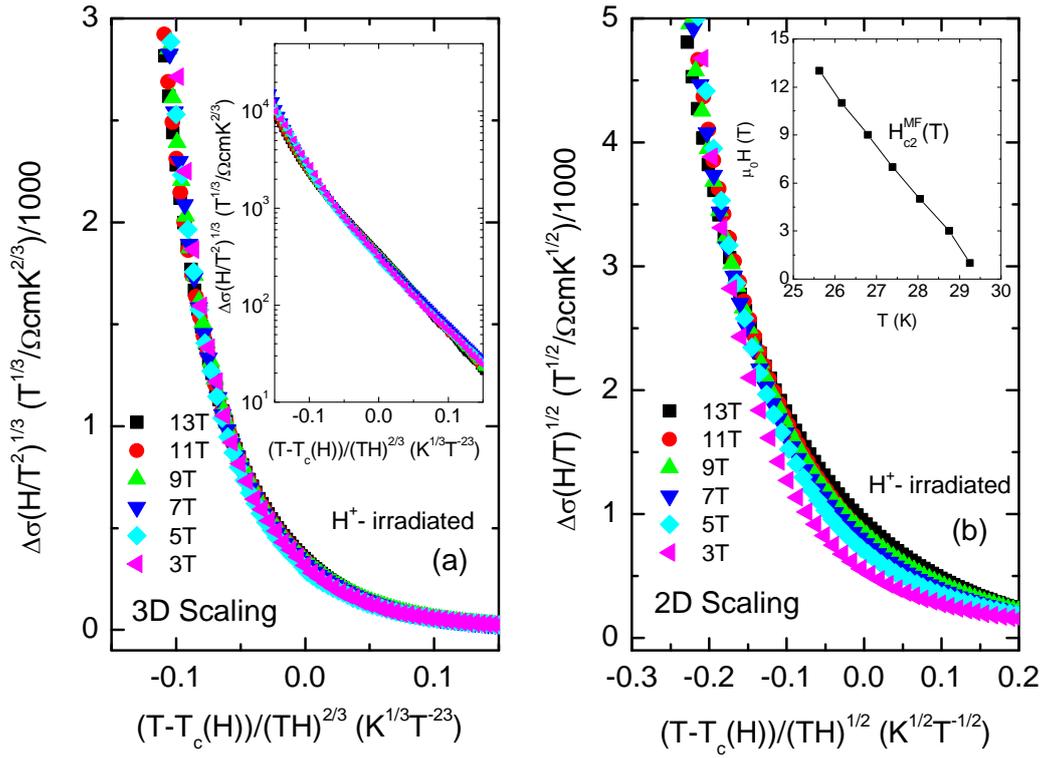

**Fig. 6.** In the main panels, the plots of the quantity $\Delta\sigma(H/T^2)^{1/3}$ vs $[T - T_c(H)]/(TH)^{2/3}$ (a) and $\Delta\sigma(H/T)^{1/2}$ vs $[T - T_c(H)]/(TH)^{1/2}$ (b) expected for the 3D UD and 2D UD cases are shown for the proton-irradiated $Ca_{8.5}La_{1.5}(Pt_3As_8)(Fe_2As_2)_5$. In the Inset of (a), the same plots are displayed in semi-logarithmic scale. In the inset (b) shows the upper critical fields $H_{c2}^{MF}$ estimated from the mean field transition temperatures.